\documentclass[journal]{IEEEtran}

\usepackage{graphicx}  

\usepackage{amsmath}   
\begin{document}
%
\title{Electron Mobility in Silicon Nanowires}
%
%
\author{E.~B.~Ramayya, D.~Vasileska,~
        S.~M.~Goodnick,~and~I.~Knezevic
\thanks{This work has been supported by the Wisconsin Alumni Research
Foundation (WARF) and Intel Corporation.}
\thanks{E. B. Ramayya and I. Knezevic are with the Department of Electrical and Computer Engineering,
University of Wisconsin -- Madison, Madison, WI 53706, USA (email:
ramayya@wisc.edu, knezevic@engr.wisc.edu). D. Vasileska and S. M.
Goodnick are with the Fulton School of Engineering, Arizona State
University, Tempe, AZ 85287, USA (email: vasileska@asu.edu,
goodnick@asu.edu).}%
\thanks{Journal reference: E. B. Ramayya, D. Vasileska, S. M. Goodnick, and I. Knezevic. IEEE Trans. Nanotech. \textbf{6}, 113 (2007).}}

%
%
%
\markboth{IEEE TRANSACTIONS ON NANOTECHNOLOGY,~Vol.~6,
No.~1,}{Ramayya~\MakeLowercase{\textit{et al.}}: Electron Mobility
in Silicon Nanowires}

%



\maketitle

\begin{abstract}
The low-field electron mobility in rectangular silicon nanowire
(SiNW) transistors was computed using a self-consistent
Poisson-Schr\"{o}dinger-Monte Carlo solver. The behavior of the
phonon-limited and surface-roughness-limited components of the
mobility was investigated by decreasing the wire width from 30 nm to
8 nm, the width range capturing a crossover between two-dimensional
(2D) and one-dimensional (1D) electron transport. The phonon-limited
mobility, which characterizes transport at low and moderate
transverse fields, is found to decrease with decreasing wire width
due to an increase in the electron-phonon wavefunction overlap. In
contrast, the mobility at very high transverse fields, which is
limited by surface roughness scattering, increases with decreasing
wire width due to volume inversion. The importance of acoustic
phonon confinement is also discussed briefly.
\end{abstract}

\begin{keywords}
Silicon nanowires, surface roughness, electron mobility
\end{keywords}

%
\IEEEpeerreviewmaketitle

\setcounter{page}{113}


\section{Introduction}
\PARstart{A}{}key factor behind the growth of the semiconductor
industry for the past 40 years has been the continuous scaling of
the device dimensions to obtain higher integration and performance
gain. According to the International Technology Roadmap for
Semiconductors (ITRS) 2005, the scaling of CMOS technology will
continue for at least another decade \cite{ITRS}. Scaling trends
also indicate that the devices of the next decade will be quasi-one
dimensional (Q1D) nanowires (NW) with spatial confinement along two
directions. Due to the compatibility with the existing fabrication
facilities, and the superior performance of silicon-on-insulator
(SOI) based metal-oxide-semiconductor field effect transistors
(MOSFETs) \cite{ColingeSOI}, ultra-narrow SOI NWs are expected to
play a critical role in future technology nodes. Therefore, it is
crucial to accurately model and estimate the performance of these
devices.

The low-field electron mobility is one of the most important
parameters that determine the performance of a field-effect
transistor. Due to the reduced density of states for scattering in
1D structures, the electron mobility in nanowires is expected to
increase significantly \cite{SakakiJJAP}. Some experimentalists
\cite{CuiNL,KooNL} claim to have observed such enhancement of
electron mobility in SiNW FETs. But Kotlyar {\it et al}.
\cite{KotlyarAPL} have shown that, in a cylindrical SiNW, the
phonon-limited mobility decreases with decreasing diameter due to an
increased overlap between the electron and phonon wavefunctions.
Recent work also shows that surface roughness scattering (SRS)
becomes less important in ultra small SiNWs \cite{WangAPL}.

In this work, we investigate the mobility of electrons in a
rectangular SiNW by taking into account electron scattering due to
acoustic phonons, intervalley non-polar optical phonons, and
imperfections at the Si-SiO$_2$ interface, and also discuss the
importance of incorporating the confinement of acoustic phonons in
these structures. Section \ref{DevStr} describes the device
structure used in this study and the components of the simulator
employed to calculate the mobility. The simulation results are
presented in Section \ref{SimRes}, while concluding remarks and a
brief summary are presented in Section \ref{Concl}.

\section{Device Structure and Mobility calculation}\label{DevStr}

\begin{figure}
\centering
\includegraphics[width=2.5in]{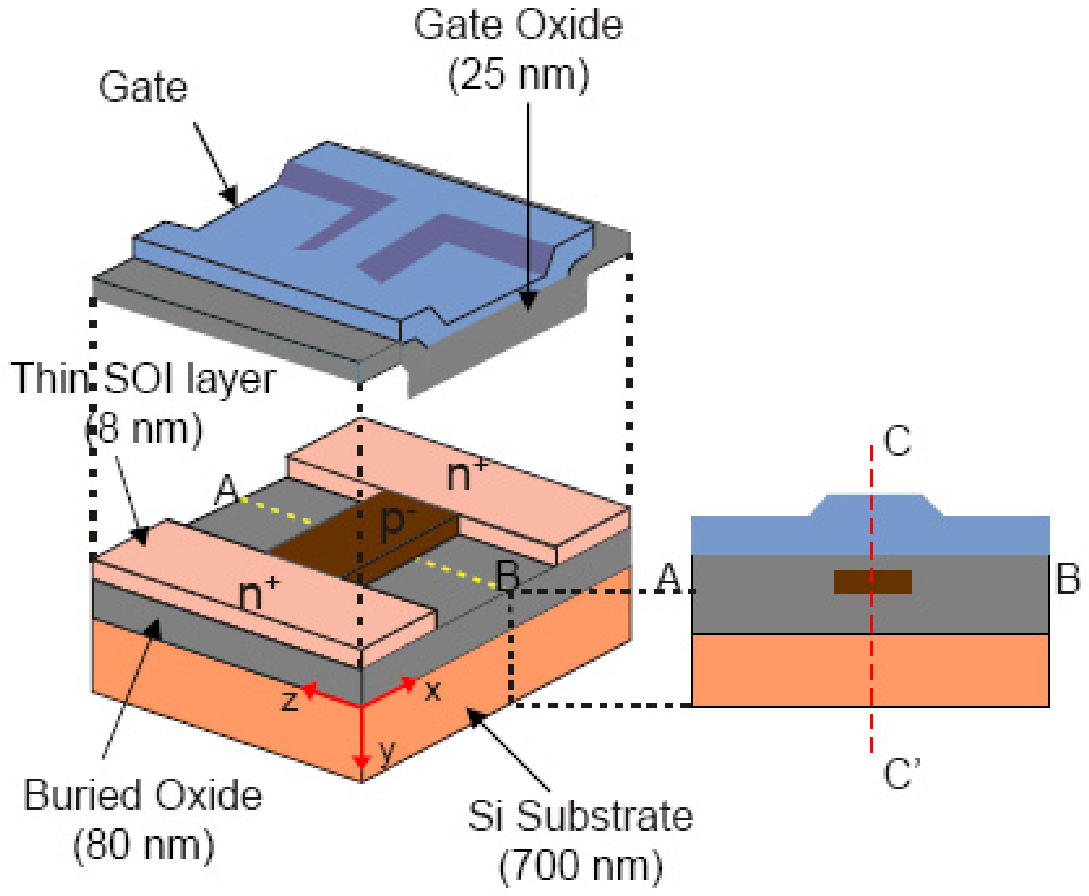}
\centering
\includegraphics[width=2.5in]{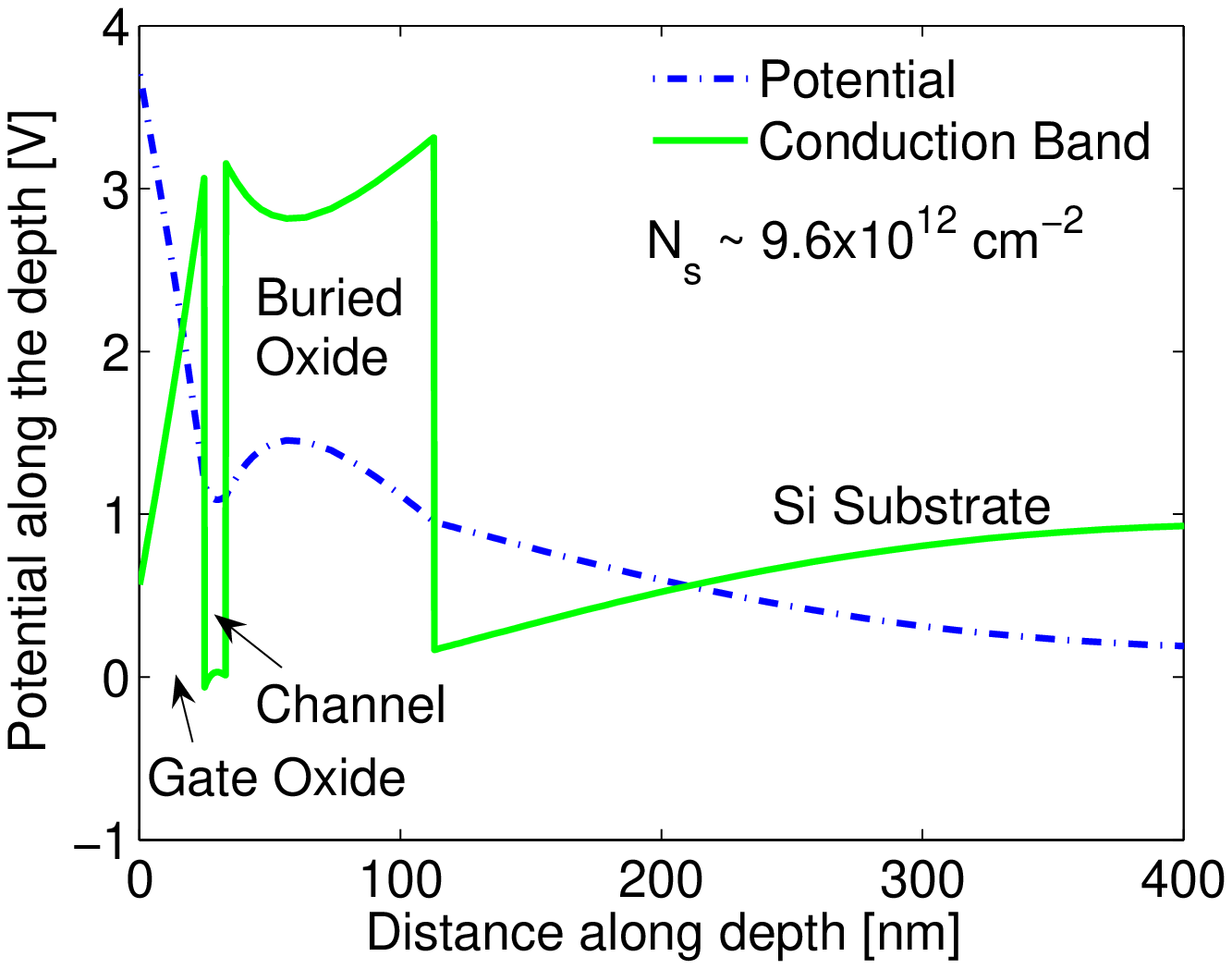}
\caption{The top panel shows the schematic of the simulated SiNW on
ultrathin SOI. The conduction band profile depicted in the bottom
panel is taken along the red cutline $CC^{\prime}$ from the top
panel. The width of the channel is 30 nm.} \label{DevStrPotY}
\end{figure}

The structure considered in this work is a long, narrow SiNW on
ultrathin SOI, similar to the channel of the device depicted in Fig.
\ref{DevStrPotY} that was originally proposed by Majima {\it et al}
\cite{MajimaEDL}. This ultra-narrow SOI MOSFET has a 700 nm silicon
substrate, an 80 nm buried oxide, an 8 nm thick SOI layer and a 25
nm gate-oxide. The width of the SiNW in the present simulation is
varied from 30 nm to 8 nm, and the channel doping is uniform at
$3\times 10^{15} cm^{-3}$. A schematic of the device and the
potential profile along the cutline $CC^\prime$, obtained by
self-consistently solving the 2D Poisson and 2D Schr\"{o}dinger
equations, are shown in Fig. \ref{DevStrPotY}. The large wire length
enables us to approximate it as infinite in the direction of the
current flow, so only the carriers' lateral momenta (and not their
positions) need to be updated in the Monte Carlo kernel. The long
wire also implies that transport is diffusive (the length exceeds
carrier mean free path), which justifies the use of semi-classical
Monte Carlo simulation.

As mentioned before, only phonon scattering and surface-roughness
scattering  are considered in this work. SRS was modeled using
Ando's model \cite{AndoRMP}, and the phonons were treated in the
bulk mode approximation. Since the wire is very lightly doped, the
effect of impurity scattering was not included. A nonparabolic band
model for silicon, with nonparabolicity factor $\alpha=0.5 eV^{-1}$,
was used in the calculation of the scattering rates. The details of
the simulator and the derivation of 1D scattering rates can be found
in Ref. \cite{RamayyaASU}; here, we present only the final
expressions used for the scattering rates calculation.

The intravalley acoustic phonon scattering rate, assuming elastic
scattering and the equipartition approximation, is given by
\begin{equation}\label{AcScatRate}
\Gamma^{ac}_{nm}(k_x)=\frac{\Xi^2_{ac}k_BT\sqrt{m^*}}{\sqrt{2}\hbar^2\rho\upsilon^2}\
\mathcal{D}_{nm}\frac{(1+2\alpha\mathcal{E}_f)}{\sqrt{\mathcal{E}_f(1+\alpha\mathcal{E}_f)}}\
\Theta(\mathcal{E}_f),
\end{equation}
where $\Xi_{ac}$ is the acoustic deformation potential, $\rho$ is
the crystal density, and $v$ is the sound velocity. $n$ and $m$ are
the initial and final subband index, respectively, while
$\mathcal{E}_n$ and $\mathcal{E}_m$ are the corresponding subband
energies. $k_x$ is the initial wavevector in the $x$-direction, in
which the motion is unconfined; $\mathcal{E}_{k_x}$ is the initial
(parabolic) kinetic energy associated with $k_x$, and
$\mathcal{E}_f$ is the final kinetic energy, given by
\begin{subequations}
\begin{eqnarray}\label{acc_energy}
\mathcal{E}_{k_x}&=&\frac{\hbar^2k^2_x}{2m^*},\\
\mathcal{E}_f&=&\mathcal{E}_n-\mathcal{E}_m+\frac{\sqrt{1+4\alpha\mathcal{E}_{k_x}}-1}{2\alpha}.\label{Ef}
\end{eqnarray}
\end{subequations}
$\mathcal{D}_{nm}$ represents the electron-phonon wavefunction
overlap  integral \cite{KotlyarAPL}, given by
\begin{equation}\label{OverlapPh}
\mathcal{D}_{nm}=\iint|\psi_n(y,z)|^2|\psi_m(y,z)|^2\,dy\,dz\, ,
\end{equation}
where $\psi_n(y,z)$ and $\psi_m(y,z)$ are the electron wavefunctions
in subbands $n$ and $m$, respectively.

The intervalley non-polar optical phonon scattering rate is given by
\begin{equation}\label{OpScatRate}
\begin{aligned}
\Gamma^{op}_{nm}(k_x)=\frac{\Xi^2_{op}\sqrt{m^*}}{2\sqrt{2}\hbar\rho\omega_0}\
&\left(N_{0}+\frac{1}{2}\pm\frac{1}{2}\right)\
\mathcal{D}_{nm} \\
&\times\frac{(1+2\alpha\mathcal{E}_f)}{\sqrt{\mathcal{E}_f(1+\alpha\mathcal{E}_f)}}\
\Theta(\mathcal{E}_f),
\end{aligned}
\end{equation}
where $\Xi_{op}$ is the optical deformation potential, and
$\mathcal{D}_{nm}$ is defined in (\ref{OverlapPh}). The
approximation of dispersionless bulk optical phonons of energy
$\hbar\omega_0$ was adopted, where
$N_{0}=\left[{\exp(\hbar\omega_0/k_BT)-1}\right]^{-1}$ is their
average number at temperature $T$. The Heaviside step function
$\Theta(\mathcal{E}_f)$ ensures the conservation of energy after
scattering, so the final kinetic energy $\mathcal{E}_f$ in the case
of optical phonons is similar to that in the acoustic phonon
expression (\ref{Ef}), but with an additional $\mp\hbar\omega_0$ on
the right-hand-side to account for the emission/absorption of a
phonon of energy $\hbar\omega_0$.

Assuming exponentially correlated surface roughness
\cite{GoodnickPRB} and incorporating the wavefunction deformation
using Ando's model \cite{AndoRMP}, the SRS rate is given by
\begin{equation}\label{SRScatRate}
\begin{aligned}
\Gamma^{sr}_{nm}(k_x,\pm)=\frac{4\sqrt{m^*}e^2}{\hbar^2}&\frac{\Delta^2\Lambda}{2+(q^{\pm}_x)^2\Lambda^2}|\mathcal{F}_{nm}|^2
\\
\times
&\frac{(1+2\alpha\mathcal{E}_f)}{\sqrt{\mathcal{E}_f(1+\alpha\mathcal{E}_f)}}\
\Theta(\mathcal{E}_f),
\end{aligned}
\end{equation}
where $\mathcal{E}_f$ is given by Eq. (\ref{Ef}), and $\Delta$ and
$\Lambda$ are the r.m.s. height and the correlation length of the
fluctuation at the Si-SiO$_2$ interface, respectively. To fit the
experimental data, $\Delta$ = 0.45 nm and $\Lambda$ = 2.5 nm are
used in this work to characterize the SRS due to each of the four
interfaces. $q_x^{\pm}=k_x-k_x'$ is the transferred wavevector,
where the angle between the initial ($k_x$) and the final ($k_x'$)
electron wavevector  can only be 0 or $\pi$, corresponding to the
$\pm$ in the superscript. The SRS overlap integral in Eq.
(\ref{SRScatRate}), for the top and bottom interface, is given by
\begin{equation}\label{OverlapSRy}
\begin{aligned}
\mathcal{F}_{nm}=&\iint
dy\,dz\left[\psi_m(y,z)\mathcal{E}_m\frac{\partial{\psi_n(y,z)}}{\partial{y}} \right. \\
&+\left.\psi_n(y,z)\varepsilon_y(y,z)\psi_m(y,z) \right.\\
&+\left.\psi_n(y,z)\mathcal{E}_n\frac{\partial{\psi_m(y,z)}}{\partial{y}}\
\right].
\end{aligned}
\end{equation}

\noindent The SRS rate and its corresponding overlap integral for
the two side interfaces can be obtained by interchanging $y$ and $z$
in Eq. (\ref{OverlapSRy}).

Using the wavefunctions and potential obtained from the
self-consistent Poisson-Schr\"{o}dinger solver, the scattering rates
are calculated. The phonon deformation potentials were taken from
Ref. \cite{TakagiJJAP}. A Monte Carlo transport kernel
\cite{JacoboniRMP} is used to model electron transport in the
unconfined $x$ direction under the influence of a very low lateral
electric field. The mobility of the electrons in the channel is then
calculated from the ensemble average of the electron velocities
\cite{JacoboniRMP}.

\section{Simulation Results}\label{SimRes}

\subsection{Validation of the Simulator}

\begin{figure}
\centering
\includegraphics[width=2.5in]{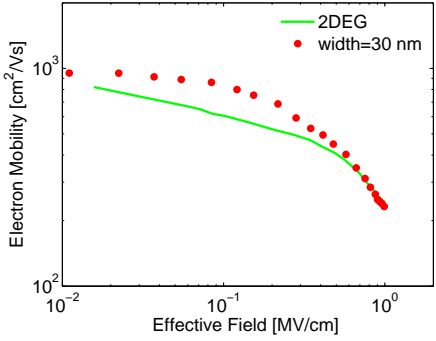}
\caption{\label{Fig2} Low-field mobility in a 30 nm wide SiNW as a
function of the effective transverse field, as obtained from the
simulation (filled circles) and the experiment of Koga {\it et al}.
\cite{KogaED} for a 2D electron gas (solid line).} \label{Mu30}
\end{figure}
A device with a width of 30 nm [at this width, electrons in the
channel feel very weak spatial confinement along the width direction
and therefore behave like a two-dimensional electron gas (2DEG)] was
used to compare the mobility results obtained from our simulator
with the experimental data of Koga {\it et al.} \cite{KogaED} for a
2DEG of the same thickness. Fig. \ref{Fig2} shows the calculated
low-field electron mobility variation with the transverse effective
field. Although there is a good agreement with the experimental data
at high fields, we find that the simulator overestimates the
mobility in the moderate and low-field regions, where phonon
scattering dominates. This discrepancy is due to the bulk phonon
approximation used in the calculation of the electron-phonon
scattering rates, and a similar result has been reported in
ultra-thin SOI structures \cite{GamizSOI,DonettiAPL}. The importance
of including phonon confinement in the calculation of the
electron-phonon scattering rates in nanostructures has previously
been established \cite{YuPRB,SvizhenkoJP}. In ultra-thin and
ultra-narrow structures, the phonon spectrum is modified due to the
mismatch of the sound velocities and dielectric constants between
the wire and the surrounding material
\cite{PokatilovSM,PokatilovPRB}, in our case -- silicon and SiO$_2$.
Presently, we are working on incorporating confined acoustic phonons
in the calculation of electron-phonon scattering rates in SiNWs, and
the results will be presented in a subsequent publication. Even for
a relatively wide, 30-nm wire, preliminary results indicate that the
acoustic phonon scattering rate is significantly higher when phonon
confinement is included in the calculation.

\subsection{Effect of Decreasing Channel Width}

\begin{figure}
\centering \centering
\includegraphics[width=2.5in]{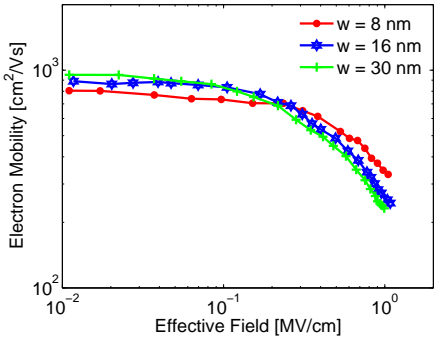}\label{Fig3}
\caption{Variation of the field-dependent mobility with varying SiNW
width. The wire thickness is kept constant at 8 nm.} \label{MuVsW}
\end{figure}
The variation of the field-dependent mobility with decreasing
channel width was investigated on a series of SiNWs, while keeping
the channel thickness at 8 nm. Fig. \ref{MuVsW} shows the mobility
for SiNWs with the widths of 30 nm, 16 nm and 8 nm. Two important
results regarding the mobility behavior in the width range
considered can be deduced from Fig. \ref{MuVsW}: (i) the mobility at
high transverse fields, which is dominated by SRS, increases with
decreasing wire width and (ii) the mobility at low-to-moderate
transverse fields, determined by phonon scattering, decreases with
decreasing wire width.

\subsubsection{Effect of Bulk Phonon Scattering}

\begin{figure}
\centering \centering
\includegraphics[width=2.5in]{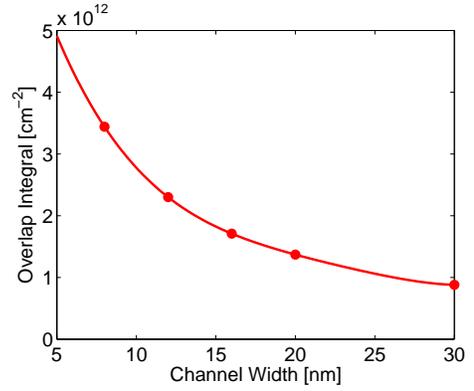}
\caption{Variation of the overlap integral for the lowest subband
with varying wire width, obtained from (\ref{OverlapPh}) at a
constant sheet density of $N_s=2.9\times10^{11} cm^{-2}$.}
\label{OverlapVsW}
\end{figure}

Phonon scattering variation with decreasing wire width is determined
by the interplay of two opposing factors: (i) reduction of the final
density of states for the electrons to scatter to, and (ii) an
increase in the electron-phonon wavefunction overlap
(\ref{OverlapPh}). The former results in an enhanced mobility, while
the latter results in mobility degradation. The overlap integral
(\ref{OverlapPh}) shown in Fig. \ref{OverlapVsW} for various widths
increases with a decrease in the wire width due to an increase in
the electron confinement. In narrow wires, the increase in the
electron-phonon wavefunction overlap dominates over the
density-of-states reduction, resulting in a net decrease in the
electron mobility at low-to-moderate transverse fields.

\begin{figure}
\centering
\includegraphics[width=2.5in]{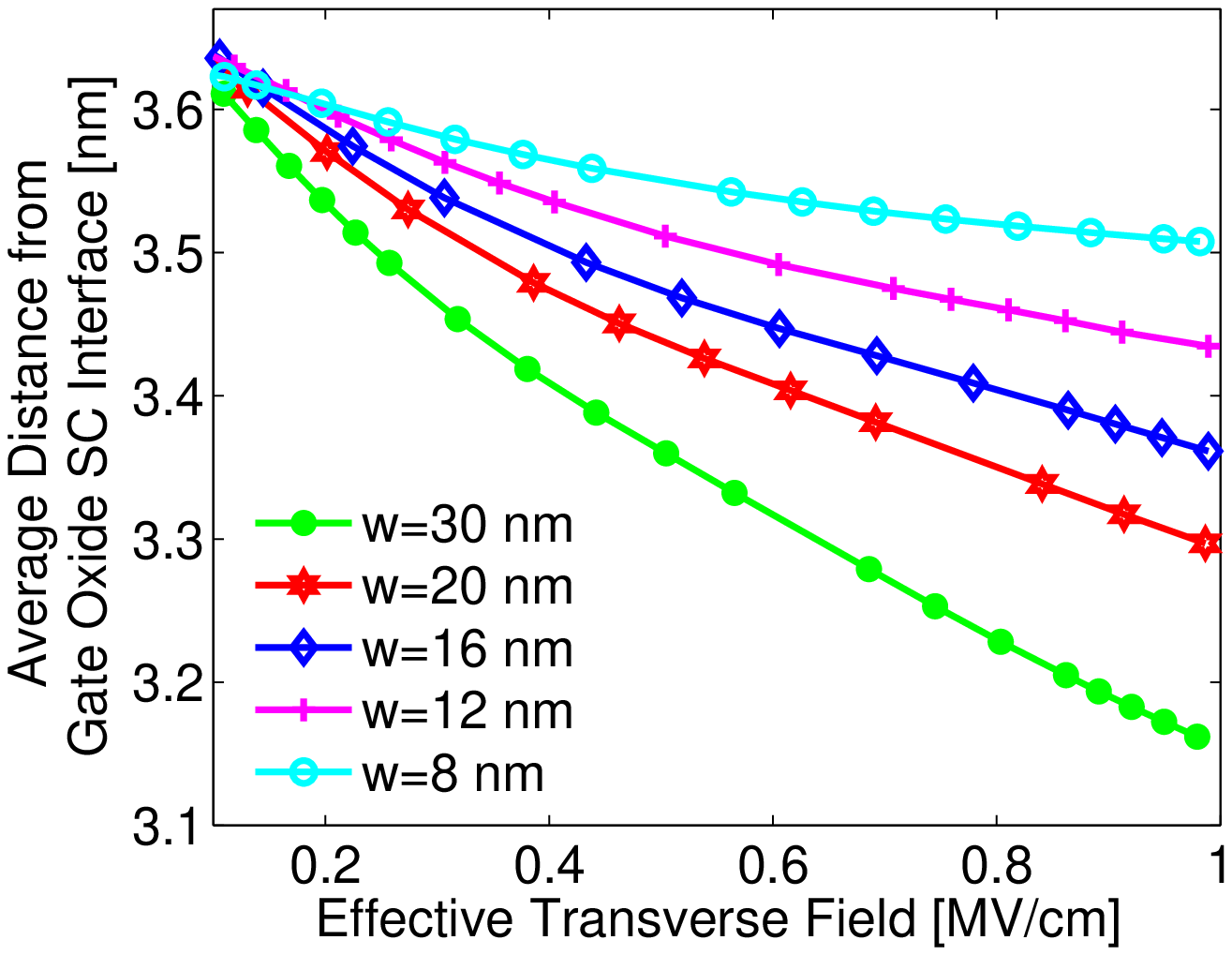}
\caption{Variation of the average distance of carriers from the top
interface (below the gate) for various wire widths as a function of
the effective transverse field.} \label{AvDist}
\includegraphics[width=2.5in]{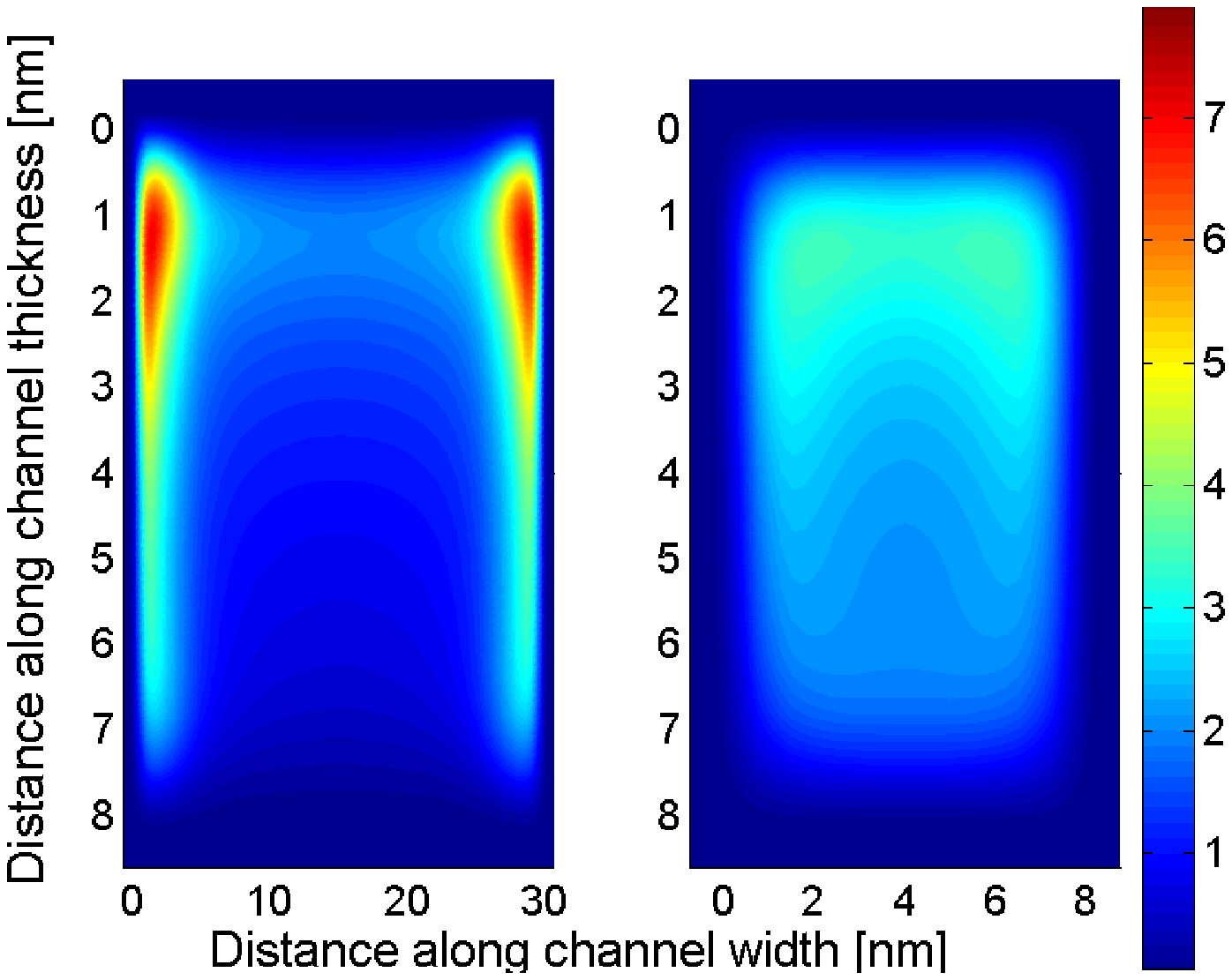}
\caption{Electron distribution across the nanowire, for the wire
width of 30 nm (left panel) and 8 nm (right panel). In both panels,
the transverse field is 1 MV/cm, the wire thickness is 8 nm, and the
color scale is in $\times10^{19} cm^{-3}$. } \label{eDen}
\end{figure}

\subsubsection{Effect of Surface Roughness Scattering}

The SRS overlap integral given by (\ref{OverlapSRy}) has three
terms. Two are due to deformation of the wavefunction, and the
third, dominant term depends on the strength of the field
perpendicular to the interface. Since the field normal to the side
interfaces is very weak, SRS due to these interfaces is much less
efficient than scattering due to the top and bottom ones. The
decrease in SRS with decreasing wire width can be understood by
following the behavior of the average distance (\ref{avDistEq}) of
the carriers from the top interface:
\begin{equation}\label{avDistEq}
\langle y\rangle
=\frac{1}{N_l}\sum_{i,\nu}N^{i,\nu}_l\iint|\psi^\nu_i(y,z)|^2 y\
\,dy\,dz\quad ,
\end{equation}
where $N_l$ is the total line density and $N^{i,\nu}_l$ is the line
density in the $i^{th}$ subband of the $\nu^{th}$ valley. In Fig.
\ref{AvDist}, we can see that the carriers are moving away from the
top interface as the width of the wire is decreased, and are
therefore not strongly influenced by the interface. This behavior is
also observed in the case of ultrathin double-gate SOI FETs, and is
due to the onset of volume inversion \cite{BalestraEDL,GamizJAP}. As
the SiNW approaches the volume inversion limit, carriers cease to be
confined to the interfaces, but are distributed throughout the
silicon volume. Fig. \ref{eDen} shows the distribution of carriers
in a 30 nm wire and a 8 nm wire at the same effective field; we can
clearly see that the carriers are confined extremely close to the
interface in the 30 wide nm wire, whereas they are distributed
throughout the silicon layer in the case of the 8 nm wire.

\section{Conclusion}\label{Concl}
The transverse-field dependence of the low-field electron mobility
in rectangular silicon nanowires was calculated using a
self-consistent 2D Schr\"{o}dinger-Poisson--1D Monte Carlo
simulation. The effects of varying the wire width and the relative
importance of phonon scattering and surface roughness scattering
were investigated. For widths in the range between 30 nm and 8 nm,
the phonon-limited mobility (dominant at low to moderate transverse
fields) was found to decrease with decreasing nanowire width,
because of the increase in the electron-phonon wavefunction overlap.
Surface roughness scattering was found to decrease with decreasing
width due to volume inversion. At high transverse fields, volume
inversion results in an appreciable mobility enhancement.

\end{document}